\documentstyle[12pt]{article}
\def\al{\alpha}
\def\be{\beta}

\def\de{\delta}

\def\et{\eta}

\def\ph{\phi}

\def\Ph{\Phi}

\def\ap{\al^\prime}

\def\ket#1{|{#1}\rangle}

\def\half{{\textstyle{1\over 2}}}
\def\frac#1#2{{\textstyle{{#1}\over {#2}}}}

\def\lsim{\mathrel{\rlap{\lower4pt\hbox{\hskip1pt$\sim$}}
    \raise1pt\hbox{$<$}}}
\def\gsim{\mathrel{\rlap{\lower4pt\hbox{\hskip1pt$\sim$}}
    \raise1pt\hbox{$>$}}}
\def\sqr#1#2{{\vcenter{\vbox{\hrule height.#2pt
         \hbox{\vrule width.#2pt height#1pt \kern#1pt
         \vrule width.#2pt}
         \hrule height.#2pt}}}}

\newcommand{\beq}{\begin{equation}}
\newcommand{\eeq}{\end{equation}}
\newcommand{\bea}{\begin{eqnarray}}
\newcommand{\eea}{\end{eqnarray}}
\newcommand{\rf}[1]{(\ref{#1})}

\renewenvironment{thebibliography}[1]
 { \rm
   \begin{list}{\arabic{enumi}.}
    {\usecounter{enumi} \setlength{\parsep}{0pt}
     \setlength{\itemsep}{3pt} \settowidth{\labelwidth}{#1.}
     \sloppy
    }}{\end{list}}

\begin{document}
\titlepage

\begin{flushright}
{IUHET 319\\}
{UATP-95/05\\}
{October 1995\\}
\end{flushright}
\vglue 1cm
	    
\begin{center}
{{\bf EXPECTATION VALUES, LORENTZ INVARIANCE, AND CPT 
\\}
\bigskip
{\bf IN THE OPEN BOSONIC STRING 
\\}
\vglue 1.0cm
{V. Alan Kosteleck\'y$^a$ and R. Potting$^b$\\} 
\bigskip
{\it $^a$Physics Department\\}
\medskip
{\it Indiana University\\}
\medskip
{\it Bloomington, IN 47405, U.S.A.\\}
\vglue 0.3cm
\bigskip
{\it $^b$U.C.E.H.\\}
\medskip
{\it Universidade do Algarve\\}
\medskip
{\it Campus de Gambelas\\}
\medskip
{\it 8000 Faro, Portugal\\}

\vglue 0.8cm
}
\vglue 0.3cm

\end{center}

{\rightskip=3pc\leftskip=3pc\noindent
The issue of spontaneous breaking of Lorentz 
and CPT invariance is studied in the open bosonic string
using a truncation scheme to saturate the
string-field action at successively higher levels.
We find strong evidence for the existence 
of extrema of the action
with nonzero expectation values for certain fields.
The Lorentz- and CPT-preserving solution 
previously suggested in the literature
is confirmed through level 12 in the action.
A family of Lorentz-breaking, CPT-preserving solutions
of the equations of motion is found to persist
and converge through level 18 in the action.
Two sequences of solutions spontaneously breaking
both Lorentz invariance and CPT are discussed.
The analysis at this level involves the analytical 
form of over 20,000 terms in the static potential.

}

\vskip 1truein
\centerline{\it Accepted for publication in Physics Letters B} 

\vfill
\newpage

\baselineskip=20pt

{\bf\noindent 1. Introduction.}
Many string theories in their critical dimension
suffer from instability
due to scalar tachyons in their spectra.
It is natural to suspect that 
these theories could be stabilized
in a different vacuum determined by
nonzero expectation values for some fields.
A relatively simple example is the open bosonic string,
for which the existence of a covariant string field theory 
\cite{ew} 
makes it possible in principle to seek solutions of
the field equations that exhibit nonzero 
expectation values for the tachyon and
perhaps other fields.
Using a truncation scheme
based on the Mandelstam notion 
\cite{sm}
of amplitude saturation,
evidence has been found
for the existence of a nontrivial solution
in which stability is restored
and modifications to the physics arise
\cite{ks1}.

When the critical dimension exceeds four,
the higher-dimensional Lorentz invariance 
must also be broken to produce a physically realistic model.
A natural stringy mechanism has been identified
that can induce spontaneous Lorentz-symmetry breaking
\cite{ks2},
an effect that is impossible
in conventional renormalizable particle gauge theories.
Under suitable circumstances,
spontaneous CPT breaking can then also occur
\cite{kp1}.
However,
to date no explicit solutions of covariant string field theory
have been presented that display spontaneous
Lorentz (and CPT) breaking.
In this letter,
we address this gap in the literature.

{\bf\noindent 2. Methodology.}
In the covariant field theory for the open bosonic string,
Fourier decomposition of the string coordinates 
allows an expansion of the string field $\Ph$
as a linear combination of particle fields
dependent on the string center-of-mass coordinate $x_0^\mu$
\cite{cstsgj}.
In terms of real particle fields and in the Siegel-Feynman gauge
\cite{ws},
the expansion begins:%
\footnote{Throughout this work except where stated,
we adopt the conventions of 
refs.\ \cite{ks1,kp1}
with the choice $g = \ap = 1$.}
\bea
\ket\Phi &=&
\biggl[ \phi (x_0)
+ A_\mu (x_0) \al_{-1}^\mu
+{1\over\sqrt 2} i B_\mu (x_0) \alpha^\mu_{-2}
+{1\over\sqrt 2} B_{\mu\nu}(x_0) \alpha_{-1}^\mu \alpha_{-1}^\nu 
+\beta_1 (x_0) b_{-1} c_{-1}
\nonumber\\ 
&&+
\quad
\ldots 
+\half i D_\mu (x_0)\al_{-4}^\mu
+\ldots 
+\de_3 (x_0)b_{-1} c_{-3} 
+\ldots \biggr]
\ket{-\half}
\quad .
\label{expansion}
\eea
In the usual vacuum,
$\ph$ is the tachyon,
$A_\mu$ is the massless vector,
$B_\mu$ and $B_{\mu\nu}$ are fields 
of squared mass one,
$D_\mu$ is a field 
of squared mass three,
and the remaining fields are auxiliary.
The first-quantized string vacuum 
is denoted $\ket{-\half}$.
The various combinations of the creation operators
$\al_{-n}^\mu$, $b_{-n}$, $c_{-n}$
acting on the vacuum span a basis in 
the string Fock space.
The normalization factors of $1/\sqrt{2}$ and $1/2$
are chosen to ensure canonical kinetic terms
in the string field action in the conventions of 
ref.\ \cite{ks1,kp1}.
By definition,
the level of a field is its mass squared
in units of $1/\ap$ as measured from
the tachyon level.
Thus, the tachyon is at level zero, 
while $A_\mu$ is at level one and the $B$ fields
are at level two.

Using this expansion,
the string field theory can be 
expressed as a particle field action 
involving an infinite number of fields
and an infinite number of cubic couplings.
We are interested in nontrivial 
(stable or unstable)
vacua of the full action,
i.e.,
solutions to the equations of motion.
Since an exact analysis remains impractical at present,
we base our analysis on a level-truncation scheme
in which the string field expansion
\rf{expansion}
is truncated at successively higher levels.
For clarity in what follows,
we say that a trunction is at level $(m,n)$ 
when all fields up to and including level $m$ appear 
in the field expansion and all terms up to 
and including total level $n$ 
are taken in the cubic interactions.
Here,
the total level number of a given term in the action 
is defined as the sum of the level numbers 
of the fields in that term. 
Further details about the level-truncation scheme
and a test in a situation where the exact answer is known
may be found in
ref.\ \cite{ks1}.

We seek static solutions of the equations of motion,
so only the static part of the cubic potential is needed.
Even with this restriction,
the length of the action grows very rapidly
as the truncation level is increased.
Under certain circumstances
where all odd-level fields have vanishing expectation values,
a further simplification can be made by
excluding odd-level fields \it a priori. \rm
Other possible simplifications
include considering only expectation values of Lorentz scalars,
or allowing only expectation values
along one direction in spacetime.
Nonetheless,
the fierce growth 
of the number of terms in the action 
with level places practical constraints
on the truncation depth to which calculations are feasible.

We use symbolic-manipulation software
to produce the analytical form of the action 
at a specified truncation level
and to derive the corresponding equations of motion.
These are simultaneous coupled polynomial equations,
one for each field appearing at the chosen truncation level.
An iterative procedure is then applied to these equations 
to obtain their roots,
each of which is a set of field-expectation values.
For low truncation levels
it is possible to find all the roots,
but typically it is necessary to perform
a search to obtain extrema of interest.

Since we are interested in solutions
that persist at arbitrarily high truncations,
it is useful to have some criteria
to dismiss unpromising candidate roots.
For instance, 
vacua with nonzero imaginary parts for any field are excluded.
We also discard roots for which the expectation value of
any field or the value of the action in the new vacuum
is significantly greater than the truncation level.
As an additional check on an otherwise acceptable solution,
we performed the following test,
referred to below as the $x$ test.
It is based on the idea that 
at any given truncation level $(m,n)$
a satisfactory solution
should be dominated by expectation values of lower-level fields.
We therefore take the relevant action at level $(m,n)$,
multiply each term at total level number $k$ by $x^k$,
and substitute the root to be tested.
The result is a polynomial in the dummy variable $x$,
with the numerical coefficient of the $k$th power of $x$ 
determining the relative contribution to the action
of the couplings at level $k$ for the given root.
A good solution is expected to display 
coefficients falling rapidly with $k$.
Also,
to avoid accidental misidentification of roots,
where feasible at each truncation level
we have followed the evolution of each root
as the coefficients of higher-truncation terms in the action
are gradually increased from zero to their full value.
The sequences of solutions we present below
are adiabatically connected in this sense.

For brevity in what follows,
we present the behavior of only those fields
shown in Eq.\ \rf{expansion}.
Also,
it is calculationally convenient to
disregard the normalization factors 
of $1/\sqrt{2}$ and $1/2$
appearing in Eq.\ \rf{expansion},
so the numerical expectation values presented below
must be appropriately scaled if they are to correspond to
canonically normalized fields in our conventions.

{\bf\noindent 3. Lorentz-invariant solution.}
In the present notation,
the truncation sequence 
for the Lorentz-invariant solution of 
ref.\ \cite{ks1}
is $(0,0)$, $(2,4)$, $(2,6)$.
At level $(2,6)$,
the nontrivial expectation values in this sequence
involve the fields
$\ph$, $\be_1$,
and% 
\footnote{Note that the normalization of $B$ we use 
in this section differs by a factor $\sqrt{52}$
from that in 
ref.\ \cite{ks1}.}
$B= \et^{\mu\nu} B_{\mu\nu}/26$.
At this truncation level,
the expectation values of
these fields and the resulting action $S$
when expressed in our dimensionless units
are%
\footnote{
This solution 
is close to the canonical unstable vacuum
in the sense that the
magnitudes of field-expectation values 
exhibit a rapid decrease with the corresponding masses.
It and the other solutions discussed below
are unlikely to be among 
the large class of exact solutions generated 
in refs.\ \cite{gth} 
at the level of string field theory 
from the unnormalizable state 
created by the identity functional.}
\beq
\ph = 1.088
\quad , \qquad 
\be_1 = 0.3804 
\quad , \qquad 
B = 0.05596
\quad , \qquad 
S = -0.1944
\quad .
\label{3a}
\eeq

We have determined the actions for the additional truncations
$(4,4)$, $(4,8)$, and $(4,12)$.
The most stringent test of convergence 
is provided by the latter, 
which involves 13 fields 
with 258 terms in the cubic part of the action.
We find excellent agreement between the results of
the $(2,6)$ and $(4,12)$ truncations.
The expectation value of the tachyon
changes by less than one percent,
and the action changes by about three percent.
The largest change,
of about eight percent,
is in the auxiliary field $\be_1$.
The percentage changes relative to the $(4,8)$ truncation
are much smaller than these.

The new expectation values for fields up to level two 
at truncation $(4,12)$ are
\beq
\ph = 1.097
\quad , \qquad 
\be_1 = 0.4113 
\quad , \qquad 
B = 0.05692
\quad , \qquad 
S = -0.2002
\quad .
\label{3b}
\eeq
Several fields at the fourth level also
acquire expectation values.
In particular,
the field $\de_3$ in Eq.\ \rf{expansion}
acquires the value $\de_3 = 0.1124$. 
As expected for a convergent solution,
the largest values at level four are clearly smaller than
the largest ones at level two.
The $x$ test is well satisfied.

Figure 1 displays the values of the field expectations
and the action as a function of maximum total level
of the cubic couplings.
Convergence of the sequence is apparent.
These results provide strong evidence for a nontrivial
Lorentz-invariant solution
in the open bosonic string.

{\bf\noindent 4. Lorentz-breaking, CPT-invariant solutions.}
The general situation with Lorentz breaking is complicated,
since the vacuum value of each component of each tensor 
could in principle have its own magnitude.
We restricted our calculations to
those special cases for which
all nontrivial tensor vacuum values are diagonal
and have equal-magnitude nonzero values 
for only a fixed subset of $d$ of the 26 possible index choices.
For brevity,
we focus our attention here on the special case $d=1$,
although in fact we have obtained the analogues 
for the cases of arbitrary $d \ne 1$
of the $d=1$ sequence of Lorentz-breaking, CPT-preserving
vacua presented in this section.%
\footnote{
Note that
the existence of one nontrivial Lorentz-breaking vacuum 
implies the existence of others.
Assuming for simplicity the euclidean formulation of the theory
in which the Lorentz symmetry is the group O(26),
there is a $d(26-d)$-parameter family
of transformations converting any given root 
into other Lorentz-breaking solutions.}

For $d=1$,
any Lorentz tensor appearing in the action can be
replaced with a single quantity representing
the possible expectation value.
We found it practicable to obtain 
the action with all fields (independent of CPT properties) 
to truncation level $(6,18)$,
where there are 20,620 independent cubic couplings
after the simplifications from the 
elimination of Lorentz indices are performed.%
\footnote{The reader interested in the details of these
cubic couplings can find their analytical form
on the World Wide Web
at http://physics.indiana.edu/$\sim$kostelec/sft.html.}
The remainder of this section specifically deals with
the CPT-preserving case.

At truncation level $(2,6)$,
a complete list of solutions reveals only
one nontrivial Lorentz-breaking and CPT-preserving case
satisfying the criteria described in section 2.
It is
\beq
\ph = 0.7131
\quad , \qquad 
\be_1 = 0.2546 
\quad , \qquad 
B = 0.03979
\quad , \qquad 
S = -0.09502
\quad .
\label{4a}
\eeq
Here,
the field $B$ is taken to be the nonvanishing
component of $B_{\mu\nu}$.
This solution is part of a sequence 
that we have followed from 
level $(0,0)$
(identical for all $d$ to the 26-dimensional solution
given in section 3)
through levels
$(2,2)$, 
$(4,4)$, 
$(2,6)$, 
$(6,6)$, 
$(8,8)$, 
$(6,12)$ 
and $(6,18)$.

Figure 2 shows selected field expectation values and the action
as a function of maximum total level 
of the cubic couplings.
Two truncations at total level six are displayed,
namely, $(2,6)$ and $(6,6)$.
It is evident that the sequence is converging.
Both the subsequences 
$(0,0)$, $(2,6)$, $(4,12)$, $(6,18)$
and 
$(0,0)$, $(2,2)$, $(4,4)$, $(6,6)$, $(8,8)$
also appear to converge to the same value.

The vacuum values of the representative
fields for the solution at truncation $(6,18)$
are:
\beq
\ph = 0.6549
\quad , \qquad 
\be_1 = 0.2506 
\quad , \qquad 
B = 0.03839
\quad , \qquad 
S = -0.08439
\quad .
\label{4b}
\eeq
The representative level-four field
acquires an expectation
$\de_3 =  0.07081$.
The $x$ test applied to this solution
gives a convincing demonstration
of rapid falloff of contributions
with field level.
At all truncation levels,
this vacuum involves
nonzero expectation values for even-level fields only.
The results in Eq.\ \rf{4b}
are therefore also valid 
at truncation level $(7,21)$,
where the contributions to the cubic couplings number in the
hundreds of thousands.
In any event,
the sequence shown
strongly suggests the persistence 
and convergence at arbitrarily high truncations
of a nontrivial Lorentz-breaking solution
in the open bosonic string.

{\bf\noindent 5. Lorentz- and CPT-breaking solutions.}
Spontaneous CPT breaking requires a nonzero expectation
value for a Lorentz tensor with an odd number of indices.
A list of all solutions at truncation level $(2,6)$
displays the existence of two independent acceptable
CPT-breaking roots,
plus their CPT images.
One is 
$$
\ph = -0.5607
\quad , \qquad 
\be_1 = -0.2479
\quad , \qquad 
B1 = \pm 1.154
\quad , \qquad 
B2 = - 0.9735
\quad ,  
$$
\beq
S = -0.1908
\quad .
\label{5a}
\eeq
Here, 
we denote by $B1$ the nonvanishing component of
the one-index field $B_\mu$ appearing in Eq.\ \rf{expansion}
and by $B2$ the nonvanishing component of the
two-index field $B_{\mu\nu}$.
The field $B1$ can have either sign,
corresponding to the two CPT conjugate solutions.
The other root at truncation level $(2,6)$ is
$$
\ph = -0.8192
\quad , \qquad 
\be_1 = -0.4104
\quad , \qquad 
B1 = \pm 1.431
\quad , \qquad 
B2 = 0.5735
\quad , 
$$
\beq
S = -0.7129
\quad .
\label{5b}
\eeq

We have obtained roots at higher truncation levels
that correspond to the above solutions.
Figure 3 shows representative expectations
and the action for the first CPT-breaking sequence,
involving the vacuum \rf{5a}.
For this first sequence five roots, 
at levels $(2,6)$, $(6,6)$, $(8,8)$, $(4,12)$, and $(6,18)$,
could be confirmed to be adiabatically connected.
The first two are so similar that the different values
of $\ph$ and $B2$
are indistinguishable on the scale used.
The figure suggests the persistence of this first 
CPT-breaking sequence at high level
and is consistent with its convergence to a finite solution
as the truncation level is further increased.
At truncation level $(6,18)$,
the values of the fields up to level two and the action
are:
$$
\ph = -1.798
\quad , \qquad 
\be_1 = -1.085
\quad , \qquad 
B1 = \pm 2.783
\quad , \qquad 
B2 = -0.7327
\quad , 
$$
\beq
S = -1.348
\quad .
\label{5c}
\eeq
The values of the representative level-four fields
are $\de_3 = -0.2676$ and $D = 0.1602$.
Here, $D$ is the nonvanishing component of the
level-four field $D_\mu$.

Figure 4 displays the analogous quantities
for the second CPT-breaking sequence,
involving the root \rf{5b}.
In this case six adiabatically connected roots
were found,
involving truncation levels
$(4,4)$, $(2,6)$, $(6,6)$, $(8,8)$, $(4,12)$, and $(6,18)$.
At truncation level $(6,18)$,
the values of the fields up to level two and the action
are:
$$
\ph = -1.892
\quad , \qquad 
\be_1 = -2.407
\quad , \qquad 
B1 = \pm 3.369
\quad , \qquad 
B2 = 1.679
\quad , 
$$
\beq
S = -2.865
\quad .
\label{5d}
\eeq
The values of the representative level-four fields
are $\de_3 = -0.09983$ and $D = -0.01003$.
The figure provides good support 
for the consistency of the second sequence. 
However,
the expectation of the level-two CPT-breaking field $B1$
appears to grow almost linearly with total level number.
This suggests that the sequence may be unbounded
in the limit of the full theory,
even if the solution exists at any finite truncation level.
In both cases
the $x$-test polynomials do display the desired
behavior of decreasing coefficients with total level number,
but the fall-off is significantly less rapid than
for the solution in section 4.
It seems that,
if either CPT-breaking sequence eventually saturates,
the rate of convergence is considerably smaller 
than for the CPT-preserving case.

{\bf\noindent 6. Comments.}
The solutions found above all have vanishing values
for odd-level fields.
\it A priori, \rm
we know of no reason why this must be so.
A separate search for CPT-breaking solutions
with nonzero expectation for the level-1 vector $A_\mu$
yielded only one candidate at or below level $(3,9)$
that passed our preliminary cuts.
However,
this solution is unsatisfactory as
it does not generate an acceptable sequence.
It also fails the $x$ test,
producing a polynomial with coefficients 
oscillating by several orders of magnitude
in successive terms.

The nontrivial extrema we have found
in the previous sections display features expected
from the general analysis suggested in refs.\ \cite{ks2,kp1}.
For example,
the mechanism for spontaneous Lorentz breaking
described in ref.\ \cite{ks2} relies here
on the appearance of terms of the form $S T\cdot T$,
where $T$ is a Lorentz tensor and $S$ is a scalar.
If a scalar with linear cubic couplings
of this type acquires a wrong-sign expectation,
the corresponding tensor is destabilized.
In the open bosonic string,
the dominant scalar is the tachyon.
As expected,
we find that the nontrivial Lorentz-invariant solutions
presented above have positive expectation values
for the tachyon,
while the Lorentz-breaking ones have negative expectation values.

In principle,
in a physically realistic string model 
the spontaneous Lorentz and CPT breaking
could extend to the physical four-dimensional spacetime.
If so,
the process must incorporate some suppression mechanism,
perhaps based on a decoupling theorem if the 
observed feature of
no CPT violation by low-level fields extends to the general case.
Since CPT invariance is a fundamental property
of particle field theories
\cite{cpt},
a spontaneous CPT breakdown of this type
could provide an experimental signature for strings
\cite{kp1,kps}.
This would occur within the framework of conventional
quantum mechanics
and is therefore distinct from the suggested possible 
CPT violations that might arise in quantum gravity
\cite{sh}.

We thank R. Bluhm and S. Samuel for useful discussions.
This work was supported in part 
by the United States Department of Energy 
under grant number DE-FG02-91ER40661
and by the 
Junta Nacional de Investiga\c c\~ao Cient\'\i fica
e Tecnol\'ogica, Portugal.

\def\plb #1 #2 #3 {Phys.\ Lett.\ B #1 (19#2) #3.}
\def\mpl #1 #2 #3 {Mod.\ Phys.\ Lett.\ A #1 (19#2) #3.}
\def\prl #1 #2 #3 {Phys.\ Rev.\ Lett.\ #1 (19#2) #3.}
\def\pr #1 #2 #3 {Phys.\ Rev.\ #1 (19#2) #3.}
\def\prd #1 #2 #3 {Phys.\ Rev.\ D #1 (19#2) #3.}
\def\npb #1 #2 #3 {Nucl.\ Phys.\ B#1 (19#2) #3.}
\def\ptp #1 #2 #3 {Prog.\ Theor.\ Phys.\ #1 (19#2) #3.}
\def\jmp #1 #2 #3 {J.\ Math.\ Phys.\ #1 (19#2) #3.}
\def\nat #1 #2 #3 {Nature #1 (19#2) #3.}
\def\prs #1 #2 #3 {Proc.\ Roy.\ Soc.\ (Lon.) A #1 (19#2) #3.}
\def\ajp #1 #2 #3 {Am.\ J.\ Phys.\ #1 (19#2) #3.}
\def\lnc #1 #2 #3 {Lett.\ Nuov.\ Cim. #1 (19#2) #3.}
\def\nc #1 #2 #3 {Nuov.\ Cim.\ A#1 (19#2) #3.}
\def\jpsj #1 #2 #3 {J.\ Phys.\ Soc.\ Japan #1 (19#2) #3.}
\def\ant #1 #2 #3 {At. Dat. Nucl. Dat. Tables #1 (19#2) #3.}
\def\nim #1 #2 #3 {Nucl.\ Instr.\ Meth.\ B#1 (19#2) #3.}

\def\plb #1 #2 #3 {Phys.\ Lett.\ B {\bf #1}, #3 (19#2)}
\def\pl #1 #2 #3 {Phys.\ Lett. {\bf #1}, #3 (19#2)}
\def\mpl #1 #2 #3 {Mod.\ Phys.\ Lett.\ A {\bf #1}, #3 (19#2)}
\def\rmp #1 #2 #3 {Rev.\ Mod.\ Phys. {\bf #1}, #3 (19#2)}
\def\prl #1 #2 #3 {Phys.\ Rev.\ Lett.\ {\bf #1}, #3 (19#2)}
\def\pr #1 #2 #3 {Phys.\ Rev.\ {\bf #1}, #3 (19#2)}
\def\prd #1 #2 #3 {Phys.\ Rev.\ D {\bf #1}, #3 (19#2)}
\def\prd #1 #2 #3 {Phys.\ Rev.\ A {\bf #1}, #3 (19#2)}
\def\npb #1 #2 #3 {Nucl.\ Phys.\ {\bf B#1}, #3 (19#2)}
\def\ptp #1 #2 #3 {Prog.\ Theor.\ Phys.\ {\bf #1}, #3 (19#2)}
\def\jmp #1 #2 #3 {J.\ Math.\ Phys.\ {\bf #1}, #3 (19#2)}
\def\nat #1 #2 #3 {Nature {\bf #1}, #3 (19#2)}
\def\prs #1 #2 #3 {Proc.\ Roy.\ Soc.\ (Lon.) {\bf A#1}, #3 (19#2)}
\def\ajp #1 #2 #3 {Am.\ J.\ Phys.\ {\bf #1}, #3 (19#2)}
\def\lnc #1 #2 #3 {Lett.\ Nuov.\ Cim. {\bf #1}, #3 (19#2)}
\def\nc #1 #2 #3 {Nuov.\ Cim.\ {\bf A#1}, #3 (19#2)}
\def\jpsj #1 #2 #3 {J.\ Phys.\ Soc.\ Japan {\bf #1}, #3 (19#2)}
\def\ant #1 #2 #3 {At. Dat. Nucl. Dat. Tables {\bf #1}, #3 (19#2)}
\def\nim #1 #2 #3 {Nucl.\ Instr.\ Meth.\ {\bf B#1}, #3 (19#2)}

\end{document}